# Excitonic Effects on Optical Absorption Spectra of Doped Graphene

*Li Yang* [†]

[†]Department of Physics, Washington University, St. Louis, Missouri, 63130, USA

E-mail: lyang@physics.wustl.edu

Abstract: We have performed first-principles calculations to study optical absorption spectra of doped graphene with many-electron effects included. Both self-energy corrections and electron-hole interactions are reduced due to the enhanced screening in doped graphene. However, self-energy corrections and excitonic effects nearly cancel each other, making the prominent optical absorption peak fixed around 4.5 eV under different doping conditions. On the other hand, an unexpected increase of the optical absorbance is observed within the infrared and visible-light frequency regime (1 ~ 3 eV). Our analysis shows that a combining effect from the band filling and electron-hole interactions results in such an enhanced excitonic effect on the optical absorption. These unique variations of the optical absorption of doped graphene are of importance to understand relevant experiments and design optoelectronic applications.

Excitonic effects are important to understand optical properties of bulk semiconductors and a wide range of nanostructures[1-3]. Graphene[4-7], a two-dimensional semimetal, has ignited significant research interests on its optical properties[8-11]. For example, enhanced excitonic effects on optical spectra of intrinsic graphene have been identified by recent first-principles calculations[12] and subsequent experiments[13, 14, 35]. However, instead of perfectly intrinsic graphene, experimental samples are usually doped by impurities, substrates, and defects, *etc.*[15-19]. In particular, recent *ab initio* studies have revealed that doping may substantially modify the electronic structure and optical response of nanostructures[20]. Therefore, it is of broad interest to understand doping effects on the optical response of graphene.

On the other hand, because of the sharp energy-momentum dispersion close to the Dirac point, experiments have shown the capability to shift the Fermi level (doping level) of graphene in a wide range (a few hundreds meV) by applying an appropriate gating voltage[21, 22]. This impressively tunable doping level provides a unique opportunity to observe how excitonic effects continuously evolve with the doping level and corresponding screening, which is of fundamental interest to understand many-electron effects in solids. Moreover, such a tunable screening effect may give hope to a new degree of freedom to tailor the optical response of graphene, which is useful for optoelectronic applications.

In this work, we employ the GW-Bethe Salpeter Equation (BSE) approach to obtain optical absorption spectra of doped graphene. The doping is realized by the rigid-band doping approach that has been applied to calculate many-electron effects of doped nanostructures[20, 23]. For a fixed Fermi level, we first obtain the electronic ground state using the density functional theory (DFT)[24, 25] within the local density approximation (LDA); then the quasiparticle energy is calculated by solving the single-particle Dyson's equation within the GW approximation[26]; finally we use the two-particle picture to solve the BSE to include *e-h* interactions and obtain optical absorption spectra[27, 28].

We observe significant variations of optical absorption spectra when changing the doping level. First, the self-energy energy correction is reduced when the doping level is increased, resulting in a red shift of the single-particle absorption spectrum. Second, the excitonic effects on the prominent absorption peak around 4.5 eV are reduced as well. Interestingly, when we increase the doping level, the reduced self-energy energy correction and excitonic effects nearly cancel each other and the final position of this prominent absorption peak is almost fixed, although the profile of the peak is slightly changed.

At the same time, we find an unexpected enhancement of the optical absorbance of doped graphene within the infrared and visible-light frequency regime (1 ~ 3 eV) after including *e-h* interactions. Our analysis shows that this enhancement of excitonic effects is not from a stronger *e-h* interaction but from the band filling effect that modifies the optical activity of resonant excitonic states near the absorption edge. As a result, unlike the usual concept that enhanced excitonic effects are always related to stronger *e-h* interactions, our calculations reveal that a reduced *e-h* interaction may enhance the variation of the low-frequency optical absorption in doped graphene.

In our calculations, structures of doped graphene are fully relaxed within the DFT/LDA. The plane-wave calculation is done in a supercell arrangement using norm-conserving pseudopotentials[29] with a 60 Ry energy cutoff. The distance between neighboring graphene sheets is set to be 1.5 nm to avoid spurious interactions. A 64 x 64 x 1 k-point grid is used to ensure converged Kohn-Sham eigenvalues and wave functions. 64 x 64 x 1 and 200 x 200 x 1 k-point grids are necessary for computing the converged self-energy and optical absorption spectra, respectively. In calculating the self-energy, we have applied the general plasmon pole model to describe the dynamical screening between electrons[25]. When solve the BSE to include excitonic effects, we have applied the static *e-h* interaction approximation in addition to the Tamm-Dancoff approximation[27, 30], which are reliable in describing excitonic effects of carbon nanotubes, silicon nanowires and intrinsic graphene[2, 12, 31, 32]. Three valence bands and three conduction bands are included to calculate optical absorption spectra up to 10 eV for the incident light polarized parallel to the graphene plane. The absorption below 0.4 eV is not shown because intraband transitions[11] and electron-phonon interactions[33] are important there and our calculation does not include these factors. As the temperature increases, these two factors will be more significant.

In this study, we focus on free standing graphene. On the other hand, substrates may introduce significant effects on many-electron interactions; metallic substrates will substantially depress electron-electron and *e-h* interactions[34] but usual dielectric substrates, *e.g.*, SiC, may only give minor modifications because of their small dielectric constants. This can be seen from the good consistence between experimental measured optical absorbance[13, 14] and previous calculation[12] on pristine graphene without including substrate effects.

We consider two doping conditions in this calculation: 0.01 and 0.02 doped electrons per unit cell. The corresponding Fermi level is shifted to around 0.50 eV and 0.72 eV, respectively, which are well accessible to realistic experimental conditions[21]. In addition, within our interested doping regime, because of the symmetry of the conduction band and valence band around the Dirac point, electron and hole dopings have similar impacts on the quasiparticle energy and low-frequency optical absorption spectra. On the other hand, because of the asymmetry of σ bands according to the Dirac point, different optical absorptions in the high-frequency regime (higher than 10 eV) shall be expected for electron and hole dopings, respectively. In this Letter, we will only focus on the electron doping case and optical absorption within the low-frequency regime (less than 8 eV).

The LDA calculated band structure of graphene around the Dirac point is presented in Figure 1. Our calculation shows that the LDA band structure is not sensitive to the doping condition. However, the shifted Fermi level will introduce a band filling effect to the optical absorption spectrum even without consider many-electron effects. As shown in Figure 1, if the Fermi level is shifted by $\Delta E$ in doped graphene, any inter-band transition with an energy less than $2\Delta E$ will be blocked. Therefore, we expect the single-particle optical absorption spectrum will be cut off by the tunable Fermi level in doped graphene. This may be useful to design optical filter devices.

We have performed GW calculations to correct the LDA band structure of doped graphene and the corresponding optical absorption spectra are plotted in Figure 2 in blue curves. As the doping level increases, the self-energy correction is reduced because of a stronger screening. This results in a red shift of the peak position as seen from Figures 2 (a) to (c). For example, in graphene with 0.02 electrons doped per unit cell, the single-particle absorption peak position is shifted from around 5.15 eV of intrinsic graphene to 4.73 eV, a 420 meV change.

The optical absorption spectra with *e-h* interactions included are presented in Figure 2 as well. Significant excitonic effects on the prominent optical absorption peak around 5 eV are exhibited, such as a substantial red shift. As the doping level increases, this red shift is reduced as seen from Figures 2 (a) to (c). This is consistent with the fact that doped free carriers enhance the screening and reduce *e-h* interactions consequently. Interestingly, the variations of the GW corrections and excitonic effects nearly cancel each other. As a result, the final position of the prominent optical absorption peak is nearly independent to the doping level. This is concluded in Figure 3 (a). Other than the peak position, we find that the doping level does slightly modify the profile of this prominent absorption peak; the line shape of this absorption peak is more symmetric with a higher doping level.

Then we turn to the low-frequency optical spectra (1 ~ 3 eV), which is of particular interest because it has been approximately regarded as a constant that is useful to measure fundamental physical parameters, such as the fine-structure constant $\alpha$ [8, 10, 11]. However, doping is usually inevitable for most experimental samples, so it is critical to know if this widely existing doping will substantially change the low-frequency optical absorbance. For intrinsic graphene, excitonic effects are negligible in this regime because of the small joint density of states (JDOS)[12]. For doped graphene, because the doping reduces *e-h* interactions, we expect an even smaller excitonic effect there. In another word, the optical absorbance with and without *e-h* interactions included shall be nearly identical to each other in the low-frequency regime.

However, we observe an unexpected enhanced excitonic effect in doped graphene in the low-frequency regime of Figure 2; the two-particle optical absorbance (the red curve) is more deviated from the single-particle one (the blue curve) when the doping level is increased! For example, we use a red dashed line to guide our eyes to the optical absorbance at 2 eV in Figure 2. We read out this absorbance under different doping conditions and conclude these data in Figure 3 (b). At 2 eV, the GW-calculated single-particle absorbance is around 0.027 and it is slightly increased to be 0.028 after including *e-h* interactions for intrinsic graphene. For graphene with 0.02 electrons doped per unit cell, the GW single-particle absorbance is 0.027 but it is increased to 0.033 after *e-h* interactions included. As a result, excitonic effects on the low-frequency absorbance are enhanced in doped graphene. This seems to be contradictory to the usual idea that doping will enhance the screening and reduce *e-h* interactions and excitonic effects.

To better understand this unique excitonic effect in doped graphene, we have to follow the approach raised by Ref. 12 to better understand the exact origin of relevant excitonic states. First, we rewrite the relevant optical transition matrix element for going from the ground state $|0\rangle$ to an exciton state $|i\rangle = \sum_k \sum_v^{hole} \sum_c^{elec} A_{vck}^i |vck\rangle$ into the form[26]

$$\langle 0|\vec{v}|i\rangle = \sum_v \sum_c \sum_k A_{vck}^i \langle vk|\vec{v}|ck\rangle = \int S_i(\omega)d\omega, \quad (1)$$

where

$$S_i(\omega) = \sum_{vck} A_{vck}^i \langle vk|\vec{v}|ck\rangle \delta[\omega - (E_{ck} - E_{vk})], \quad (2)$$

which gives a measure of the contribution of all inter-band pairs $(ck, vk)$ at a given transition energy $\omega$ to the optical strength of the exciton state $i$.

As the first step, we start from those bright excitonic states around the prominent optical absorption peak because they exhibit normal behaviors, *i.e.*, increasing the doping level decreases excitonic effects. In Figure 4, $S_i(\omega)$ and its integrated value up to a given frequency are presented for two optically bright excited states around the prominent absorption peak but under different doping levels. There are some common features that we can read from Figure 4: (1) because of the resonant nature, both excitonic states take oscillator strength from inter-band transitions near them; (2) $S_i(\omega)$ displays a nodal structure across its eigenvalue; optical contributions from lower and higher energy always have opposite signs to each other; (3) the sum of higher-energy contributions is usually larger than that of lower-energy contributions so that the final integrated optical activity has the same sign as that of higher-energy contributions. These features reflect the attractive nature of *e-h* interactions and explain why excitonic effects usually introduce a red shift to the optical absorption spectra.

With the above features in our mind, we can get a better idea of excitonic states in doped graphene. For doped graphene, the screening is enhanced and *e-h* interactions are reduced so that the resonant energy range of excitonic state in Figure 4 (b) is smaller than that of intrinsic graphene in Figure 4 (a). As a result, the prominent peak of doped graphene shows a reduced excitonic effect, *e.g.*, a smaller red shift from the single-particle result, which is consistent with what we observe in Figure 2.

In Figure 5, the same plots are presented for two bright excitonic states within the low-frequency regime (around 1.5 eV) of intrinsic and doped graphene, respectively. The similar shrink of the resonant energy range is observed for the excitonic state (Figure 5 (b)) in doped graphene, showing that *e-h* interactions are reduced because of a stronger metallic screening. However, the unique feature of resonant excitonic states around the absorption edge in doped graphene is that their resonant energy range touches the cutoff energy induced by the band filling. Therefore, the negative contributions from the lower energy regime are terminated at the cutoff energy marked by a cyan arrow in Figure 5 (b). Since the total optical strength of a resonant exciton is the sum of all inter-band contributions according to Eq. (1), those blocked negative elements give rise to a stronger integrated optical strength of the excitonic state plotted in Figure 5 (b). In a word, with the help of even reduced *e-h* interactions, the band filling effect introduces an enhanced variation on the optical absorbance close to the absorption edge in doped graphene. This unique variation of the optical absorbance of doped graphene induced by excitonic effects may be of importance for relevant experiments trying to measure the fundamental physical constant through the optical absorbance of graphene.

In conclusion, we have performed first-principles calculations to study many-electron effects on the optical absorption spectra of doped graphene. Doped free carriers do enhance the screening and reduce the *e-h* interactions consequently. As a result, the red shift of the prominent peak around 4.5 eV is reduced. However, we observe an unexpected enhancement of the optical absorbance at the low-frequency regime (1 ~ 3 eV). Through our analysis, we identify that this enhanced excitonic effect is not mainly from the change of screening but from the band filling effect induced by the shifted Fermi level. Our calculated results are not only useful to understand the general picture of excitonic effects in doped semiconductors and semimetals but also of practical importance to understand relevant experimental measurements of graphene.

Figures and captions:

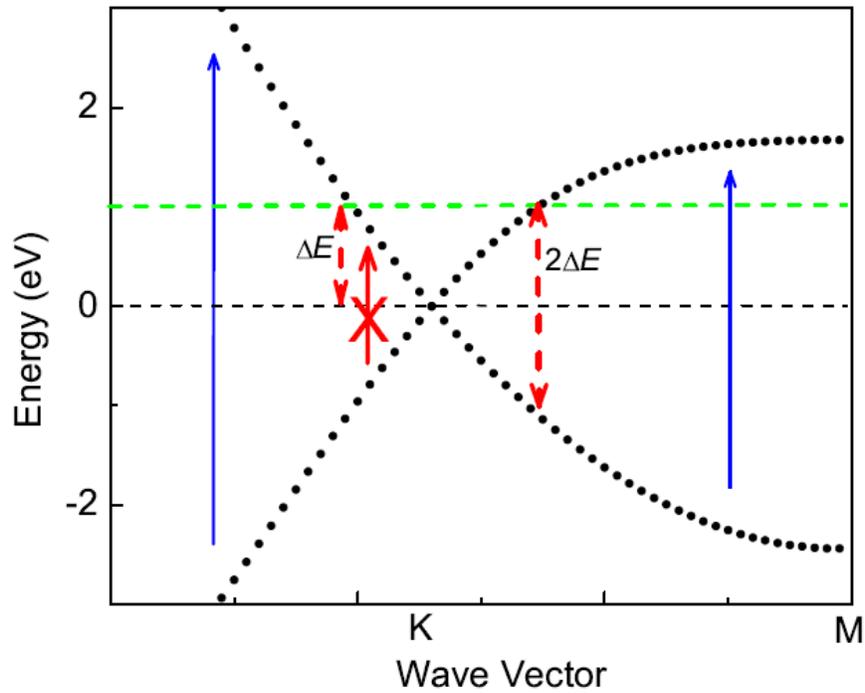

Figure 1: (Color online) LDA-calculated band structure of graphene. The doping level is represented by the dashed green line. The solid red arrows represent those blocked inter-band transitions and the solid blue arrows represent those survived transitions.

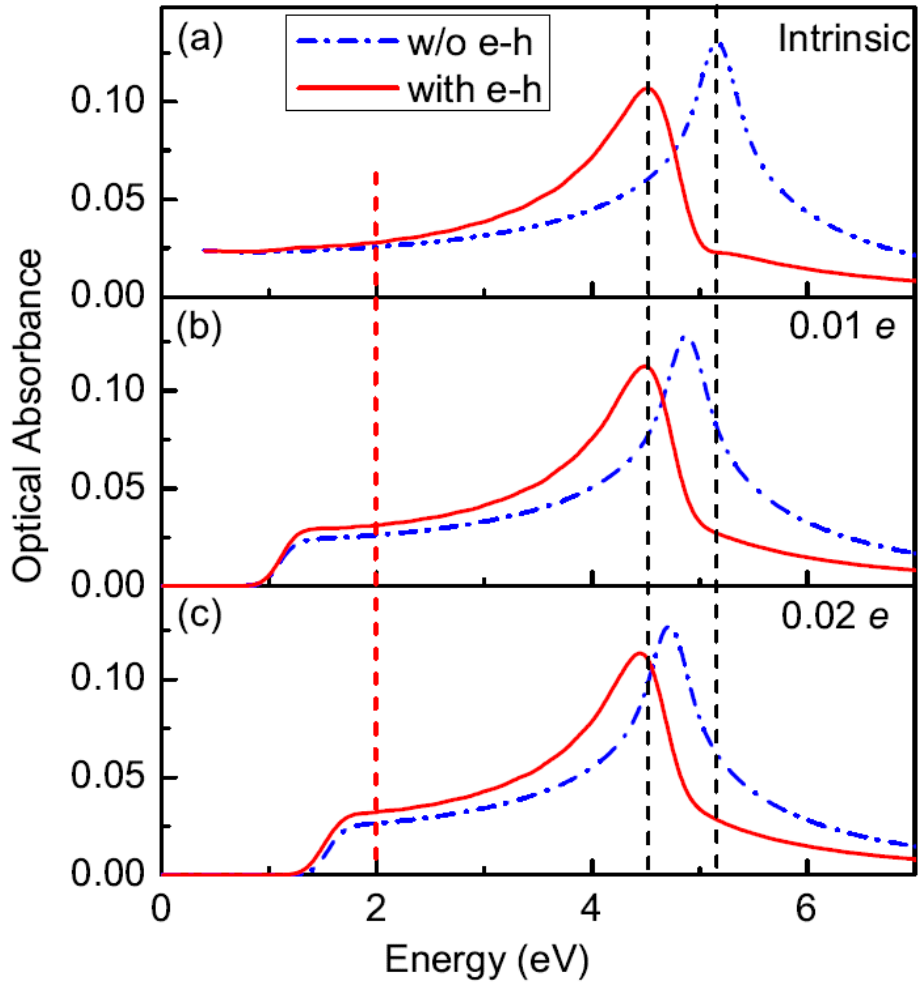

Figure 2: (Color online) Optical absorption spectra of intrinsic and doped graphene. (a) is the that of intrinsic graphene, (b) is that of doped graphene with 0.01 electron per unit cell, and (c) is that of doped graphene with 0.02 electron per unit cell. A 0.12 eV Gaussian smearing is applied to all optical absorption spectra. The vertical black and red lines are used to guide readers' eyes.

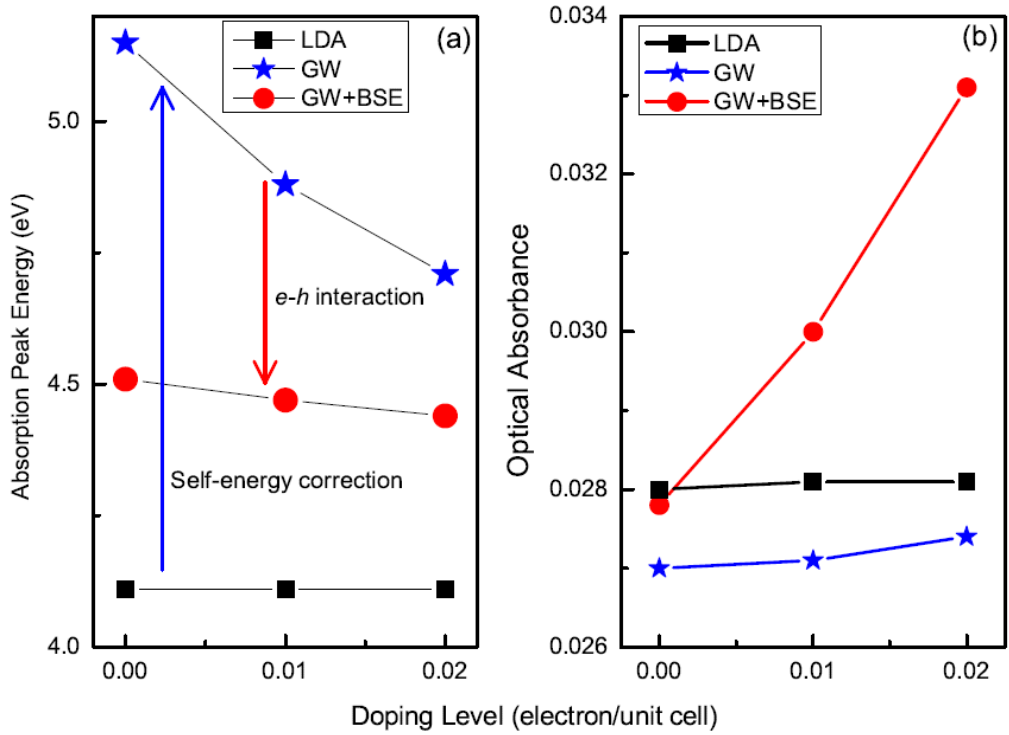

Figure 3: (Color online) (a) The position of the prominent absorption peak and (b) the optical absorbance of graphene at 2 eV under different doping conditions.

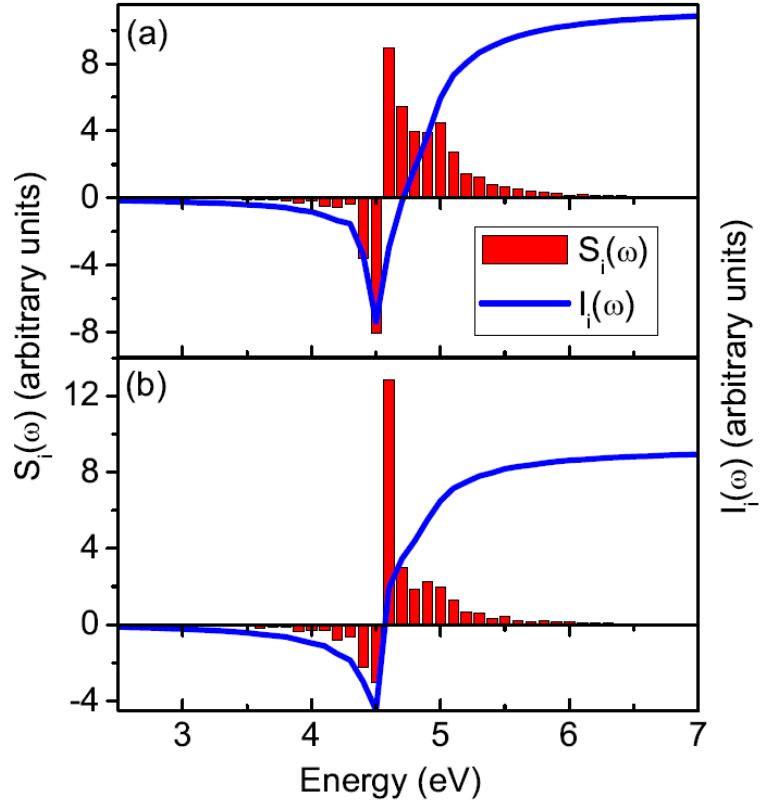

Figure 4: (Color online) $S_i(\omega)$ and the corresponding integration $I_i(\omega) = \int_0^{\omega} S_i(\omega')d\omega'$ of two optically bright states around the prominent optical absorption peak of intrinsic and doped graphene (0.01 electrons per unit cell). The eigenvalues of these two excitonic states are 4.65 eV (a) and 4.62 eV (b), respectively.

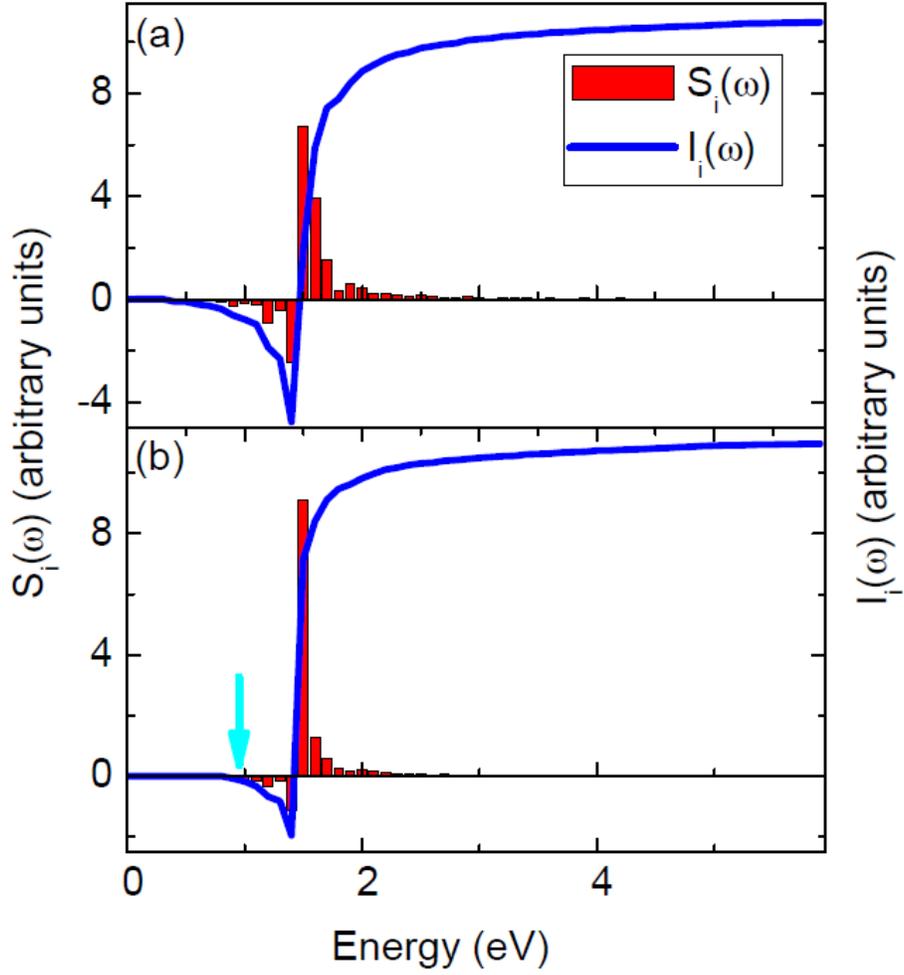

Figure 5: (Color online) $S_i(\omega)$ and the corresponding integration $I_i(\omega)$ of two optically bright states within the low-frequency regime of intrinsic and doped graphene (0.01 electrons per unit cell). The eigenvalues of these two excitonic states are 1.51 eV (a) and 1.49 eV (b), respectively. The cut-off energy by the band filling is marked by a cyan arrow in (b).